# Dynamics of Ternary Cu-Fe-S$_2$ Nanoparticles Stabilized by Organic Ligands

J. Żukrowski[1], A. Błachowski[2], K. Komędera[2], K. Ruebenbauer[2], G. Gabka[3], P. Bujak[3], A. Pron[3], and M. Przybylski[1,4*]

[1]Academic Centre for Materials and Nanotechnology, AGH University of Science and Technology
*PL-30-059 Kraków, Av. A. Mickiewicza 30, Poland*

[2]Mössbauer Spectroscopy Division, Institute of Physics, Pedagogical University
*PL-30-084 Kraków, ul. Podchorążych 2, Poland*

[3]Faculty of Chemistry, Warsaw University of Technology
*PL-00-664 Warsaw, ul. Noakowskiego 3, Poland*

[4]Faculty of Physics and Applied Computer Science, AGH University of Science and Technology
*PL-30-059 Kraków, Av. A. Mickiewicza 30, Poland*

[*]**Corresponding author:** marprzyb@agh.edu.pl

**Abstract**

Chalcopyrite-type (Cu-Fe-S$_2$) ternary nanocrystals stabilized by long aliphatic chain ligands could be considered as isolated hard nano-objects dispersed in soft network of organic ligands. The main attention was paid to the behavior of the particles whose average size was varied in a controllable manner from 3 to 20 nm. Dynamics of nanoparticles was studied by applying Mössbauer spectroscopy. The fast dynamics could be described by two-level environment. Deeper level (atomic) was practically the same as for bulk material except Debye temperature, but the higher level (particle motion) was described by the classical harmonic oscillator with the spring constant dramatically softening with increasing temperature. Such behavior led to fast decrease of the fraction detectable by Mössbauer spectroscopy with increasing temperature. The induced internal oscillations within particle by surrounding thermal bath additionally contribute to the shift of measured spectra. Slow dynamics was characterized by the thermally driven overdamped harmonic oscillator motions. In addition, the long range-like diffusion of particles was seen. No significant rotation of particles was found within the accessible temperature range.

## 1. Introduction

Nanoparticles are interesting by themselves as they provide insight into intermediate level physics between atomic and macroscopic scales[1-4]. They find many practical applications as well[5], e.g., as carriers of some molecules (including biologically active ones) in bioimaging[6,7] and drug delivery[8]. They are also used in electronics and energy conversion as components of photodiodes[9], photovoltaic cells[10] and as thermoelectric materials of new generation[11].

However they tend to cluster into larger objects due to the interactions between their surfaces. This feature is particularly enhanced for particles with internal magnetic order[12]. Stericly stabilized semiconductor nanoparticles, upon deposition on a substrate, form systems consisting of isolated hard particles (inorganic cores) embedded in a soft network of stabilizing ligands containing long aliphatic chains[4]. Dynamics of theses layers of hard/soft nature seems especially interesting, while seen from the atomic position within the particle, and in the case of semiconductor nanocrystals/organic ligands system essentially unexplored[13,14]. In this case, the first lowest atomic level is described by atomic motion within crystal forming nanoparticle, while the higher second level is described by the motion of the nanoparticle in the binder. Hence, investigation of the iron containing particles by means of the transmission Mössbauer spectroscopy is somewhat challenging and it can return interesting results. Many level dynamics has been already studied in detail for large biomolecules containing iron. These systems are somewhat different as they contain soft matter in the soft mater. In the case of nanoparticles one has hard particles embedded in soft matrix[15]. Some dispersions of nanoparticles in organic matter have been already studied by Mössbauer spectroscopy, but usually many problems with the agglomeration of the particles were encountered[16]. Hence, it is important to find reasonably stable systems, but being soft enough to see distinct dynamics of the binding medium. The papers reporting properties of such system – particularly with FePt particles have been published, but dynamics was usually not studied in detail[13,14].

Here we report the results of Mössbauer effect investigations for inorganic/organic layers consisting of ternary, chalcopyrite-type $Cu-Fe-S_2$ nanocrystals dispersed in a soft network of organic ligands containing long aliphatic chains. The prepared batches of nanocrystals exhibited the same composition ($Cu_{1.62}Fe_{1.00}S_{2.00}$), but differed in the average nanocrystal's size from 3 nm to 20 nm[4].

## 2. Experimental

In a typical synthesis of $Cu-Fe-S_2$ nanocrystals, 60 mg (0.61 mmol) of CuCl, 100 mg (0.61 mmol) of $FeCl_3$, 93 mg (1.22 mmol) of thiourea, 361 mg (2.10 mmol) of oleic acid and 15 mL of oleylamine were added to a 25 mL three-neck flask. The amounts of the reagents used in all preparation are listed in Table S1 (Supporting Information).The mixture was heated under argon flow to 120°C until a homogenous solution was formed. Subsequently the temperature was raised to 180°C, and the mixture was kept at this temperature for 60 min. Upon heating, the color changed rapidly from yellow through brown and finally to black. The mixture was then cooled to room temperature and subsequently toluene (10 mL) was added. In the next step the reaction mixture was centrifuged and the isolated black precipitate was separated. The supernatant was treated with 30 mL of acetone leading to the precipitation of the desired fraction of nanocrystals. The nanocrystals were separated by centrifugation (7000 rpm for 5 min) and re-dispersed in chloroform (alternatively in toluene or methylene chloride). Three batches of copper rich nanocrystals ($Cu_{1.62}Fe_{1.00}S_{2.00}$) of the following average size were prepared: 3 nm, 9 nm and 20 nm. For more details, see[4].

Mössbauer absorbers made of nanoparticles stabilized by organic ligands were prepared in the following way. The filter paper was soaked with chloroform suspension of particles containing organic binder and the chloroform was allowed to evaporate to dryness. Sheets of the resulting paper were stacked together to make absorber of the acceptable resonant thickness. The estimated mass of the samples used for Mossbauer experiment is 30-40 mg/cm$^2$.

Mössbauer spectra were obtained by using mainly MsAa spectrometers equipped typically with the Kr-filled proportional detectors. A commercial $^{57}$Co(Rh) source obtained from Ritverc GmbH of the nominal activity 50 mCi was used. The source was kept at room temperature, while the absorber temperature was controlled with accuracy better than 0.01 K. The velocity scale was typically calibrated by means of the Michelson interferometer equipped with the He-Ne laser[17].

Data were processed and spectra fitted using applications from the Mosgraf-2009 suite[18]. All shifts are reported versus shift in room temperature α-Fe.

## 3. Dynamics of nanoparticles

### 3.1. Fast dynamics in crystals

We arbitrary separate fast dynamics from slow dynamics due to the experimental technique applied. Fast dynamics for the 14.4-keV resonant transition in $^{57}$Fe nuclei can be defined as dynamics with the characteristic time scale being shorter than about 100 ps. For a transmission Mössbauer spectroscopy with the absorber material being investigated it affects the second order Doppler shift (SOD) and the recoilless fraction as seen on the resonant atom. The SOD comes together with the isomer shift and it is measured relative to some standard – usually source or reference material. The recoilless fraction enters dimensionless absorber thickness and usually it is measured relative to some reference temperature recoilless fraction, i.e., as the ratio $f^{(c)}/f_0^{(c)}$, where the symbol $0 < f^{(c)} < 1$ denotes (crystalline) recoilless fraction at some temperature $T$, while the symbol $0 < f_0^{(c)} < 1$ denotes corresponding recoilless fraction at the reference temperature $T_0$ provided all remaining conditions are the same. Fast dynamics is observed for a system being at thermal equilibrium[19].

*3.1.1. Second order Doppler shift (SOD)*

The SOD is defined as $\delta_D = -\langle v^2 \rangle/(2c)$ with the symbol $\langle v^2 \rangle$ denoting mean squared velocity of the resonant atom in the laboratory frame and the symbol $c$ standing for the speed of light in vacuum. For a harmonic approximation of the crystalline matter including immobilized nanoparticles one can write [20]:

$$\langle v^2 \rangle = \left(\frac{3k_B}{2m}\right) \int_0^{+\infty} dx\, x \left[\frac{\exp(x/T)+1}{\exp(x/T)-1}\right] D(x).$$

(1)

The symbol $m$ stands for the rest mass of the resonant atom in the ground state, while the symbol $k_B$ denotes Boltzmann constant. The function $D(x)$ stands for the phonon density of states (DOS) projected on the resonant atom. It is quite often approximated by the Debye function $D(x) = 3x^2 \theta_D^{-3}$ for $0 \leq x \leq \theta_D$ and $D(x) \equiv 0$ otherwise. The parameter $\theta_D > 0$ being constant in the harmonic approximation is called Debye temperature. One has to note that expression (1) approaches quite rapidly a classical non-relativistic limit $\langle v^2 \rangle = (3k_B T)/m$ with increasing temperature and becomes completely insensitive to the solid state environment of the resonant atom.

In the presence of the higher vibrational level the mean squared velocity of the resonant atom has to be replaced by $\langle v^2 \rangle_T = \langle v^2 \rangle + \langle V^2 \rangle$ provided assumptions listed above hold and higher level vibrations remain uncorrelated to the atomic vibrations. Additional mean squared velocity calculated in the classical non-relativistic limit amounts to $\langle V^2 \rangle = [3k_B(T-T_D)]/M$ for $T > T_D > 0$, and otherwise one has $\langle V^2 \rangle \equiv 0$. A temperature $T_D$ and effective vibrating mass $M$ are adjustable parameters. Substitution of the average rest mass of the even very small nanoparticle leads effectively to $\langle V^2 \rangle \approx 0$, as the term $-\langle V^2 \rangle/(2c)$ is below observation limit. Hence, this term could be observed provided some internal vibrations of the nanoparticle are generated by the surrounding thermal bath at sufficiently high temperature. Vibrations of the sufficient velocity to be observed can occur only in the "hard" nanoparticle embedded in the "soft" matrix. Hence, they were unobserved in large biomolecules with many levels of fast dynamics. One can conclude that SOD may become "less boring" at high temperature provided one has at least two levels of fast dynamics.

*3.1.2. Recoilless fraction*

For isotropic recoilless fraction described in a harmonic approximation one can use the following expression $f^{(c)} = \exp[-q^2 \langle R^2 \rangle]$, where the symbol $q = E_0/(\hbar c)$ is a very good approximation of the wave number transfer to the system with $E_0$ being resonant transition energy, and $\hbar$ the Planck constant divided by $2\pi$. The symbol $\langle R^2 \rangle$ stands for the dynamic mean squared displacement of the resonant atom in arbitrary direction and from the equilibrium position. The mean squared displacement takes on the following form for a crystalline solid including immobilized nanoparticle[20]:

$$\langle R^2 \rangle = \left( \frac{\hbar^2}{2mk_B} \right) \int_0^{+\infty} dx\, x^{-1} \left[ \frac{\exp(x/T)+1}{\exp(x/T)-1} \right] D(x).$$

(2)

One can again use Debye temperature applying approximation $D(x) = 3x^2 \theta_F^{-3}$ for $0 \leq x \leq \theta_F$ and $D(x) \equiv 0$ otherwise. The parameter $\theta_F > 0$ being constant in the harmonic approximation is called also Debye temperature. Due to the details of the real DOS Debye temperatures $\theta_F$ and $\theta_D$ are usually slightly different, but in general one can use the following approximation $\theta_D \approx \theta_F$.

One has to observe that for recoilless absorption within stationary in the laboratory frame and unsuspended nanoparticle recoil energy amounts to $E_R = E_0^2/(2M_P c^2)$, where the symbol $M_P$ stands for the rest mass of the particle. Hence, in order to get resonant absorption one needs the following condition to be satisfied $E_0/(2M_P c) < \gamma_0$, where $\gamma_0$ denotes width of the transition. One needs $M_P > 10^9$ a.u. to neglect this kind of recoil for a transition considered here, as $\gamma_0 = 0.097$ mm/s, while $E_0/(2M_P c) = 2.3 \times 10^{-3}$ mm/s for $M_P = 10^9$ a.u.. Effect described above is going to shift positively spectrum for a particle being an absorber[21,22]. However this criterion is not very relevant as particles are always bound to some ligands or between themselves except some unusual systems.

For a hard nanoparticle embedded in some soft matrix one can expect another level of the fast dynamics in addition to the described above. Namely, some additional dynamic mean squared displacement can occur due to the motion of the whole particle and excitation of some internal vibrations caused by the interaction with the thermal bath generated by the embedding medium. One can assume that this extra dynamics is classical and remains uncorrelated with the dynamics described above. One can assume that the system remains at thermal equilibrium at this higher level as well[13,14]. For such case the ratio $f^{(c)}/f_0^{(c)}$ has to be replaced by the ratio $f/f_0 = (f^{(c)} F)/(f_0^{(c)} F_0)$, where the symbols $F$ and $F_0$ stand for the respective recoilless fractions at the higher level(s). Hence, one obtains $-q^{-2} \ln(F/F_0) = \langle x^2 \rangle - \langle x^2 \rangle_0$ provided higher level dynamics is harmonic and isotropic. Symbols $\langle x^2 \rangle$ and $\langle x^2 \rangle_0$ describe additional dynamic mean squared displacements of the resonant atom in the arbitrary direction, respectively. The function $\langle x^2 \rangle - \langle x^2 \rangle_0$ could be approximated by the following expression provided the reference temperature $T_0$ is the lowest temperature within the data set:

$$\langle x^2 \rangle - \langle x^2 \rangle_0 = \alpha\, z + \tfrac{1}{2} \beta\, z^2 + \sum_{n=3}^{+\infty} \beta_n z^n \text{ with } z = T - T_0 \text{ and } \beta_{n+1} = \beta_n \left( \frac{\varepsilon}{n+1} \right) \text{ with } n \geq 3, \text{ where}$$
$$\beta_3 = \tfrac{1}{6} \beta \varepsilon.$$

(3)

Parameters $\alpha$, $\beta$ and $\varepsilon$ are adjustable parameters. Under assumption that the classical equipartition principle holds and under assumption that the vibrating mass remains constant one can obtain spring constant of the effective equivalent isotropic harmonic oscillator as:

$$K(z) = \left( \alpha + \tfrac{1}{2} \beta z + \sum_{n=3}^{+\infty} \beta_n z^{n-1} \right)^{-1}.$$

(4)

The spring constant remains constant versus temperature provided $\beta = 0$.

*3.2. Slow dynamics*

*3.2.1. Particle motions*

The primary manifestation of the slow dynamics is a motion of the particle as a whole within embedding matrix. For dense and highly viscous matrix one can exclude Brownian motion and describe motion by the jump diffusion mechanisms leading to the phase modulation of the absorbed radiation instead to the Doppler modulation being primary effect for the Brownian motion at the space-time scales considered[23]. Under assumption that the motion is locally isotropic one can distinguish two major mechanisms for the absorption line broadening. The first one is a motion described by the thermally driven overdamped isotropic classical harmonic oscillator[24], while the other one is long range like isotropic jump-like diffusion[25,26]. One can consider rotational motions as well, the latter leading to the relaxation of the eventual electric quadrupole interaction. For fast unrestricted rotations the electric quadrupole interactions are likely to average almost to null.

Motion due to the overdamped classical harmonic oscillator could be described in the simplest approximation as splitting of each absorption line (here Lorentzian) into two components having the same position, but differing by areas and linewidths. Namely, one obtains the following expressions for the respective line shapes of the narrow $L_1(\omega)$ and broad components $L_2(\omega)$ versus ambient velocity $\omega$ [24]:

$$L_1(\omega) = [(1 - C_2)(\tfrac{1}{2}\Gamma_1)^2] / [(\tfrac{1}{2}\Gamma_1)^2 + (\omega - \omega_0)^2]$$
$$L_2(\omega) = [C_2(\Gamma_1 / \Gamma_2)(\tfrac{1}{2}\Gamma_2)^2] / [(\tfrac{1}{2}\Gamma_2)^2 + (\omega - \omega_0)^2]$$
$$C_2 = b / (1 + b) \; : \; \Delta\Gamma = \Gamma_2 - \Gamma_1 = S_0 / (q\tau_R)$$

(5)

Here the symbol $\Gamma_1$ stands for the linewidth of the narrow component, while the symbol $\Gamma_2$ denotes corresponding linewidth of the broad component. The symbol $\omega_0$ stands for the common position of above lines. The parameters $b$ and $S_0$ take on the forms:

$$b = \sum_{n=1}^{+\infty} \left( \frac{(q^2 \langle \xi^2 \rangle)^n}{n!} \right) : \langle \xi^2 \rangle = \tfrac{1}{3} R^2 \text{ and } S_0 = \left( \frac{2}{\exp[q^2 \langle \xi^2 \rangle] - 1} \right) \sum_{n=1}^{+\infty} \left( \frac{(q^2 \langle \xi^2 \rangle)^n}{(n-1)!} \right).$$

(6)

Hence, one obtains two adjustable parameters $R$ and $\tau_R$ representing characteristic "radius" of the oscillator and characteristic time scale of the motion, respectively. The radius increases usually linearly with increasing temperature due to the expansion of the embedding medium, while the characteristic time scale remains fairly constant once this type of motion is started at sufficiently high temperature provided other thermodynamic parameters remain unchanged.

The long range like isotropic jump-like diffusion is likely to proceed via uncorrelated events (jumps) leading to the homogeneous broadening of the lines described above. Hence, one can write down the following expressions under assumption that diffusion remains uncorrelated to the overdamped oscillator motion[25]:

$$\Gamma_1 = \Gamma_0 + \left( \frac{2\hbar c w_D [1 - \alpha(q)]}{E_0} \right) \exp[-(U/T)] : \alpha(q) = q^{-1} \int_0^{+\infty} dr\, r\, \rho(r) \sin(qr) :$$
$$\rho(r) \geq 0 : \int_0^{+\infty} dr\, r^2\, \rho(r) = 1 : \left( \frac{2\hbar c w_D [1 - \alpha(q)]}{E_0} \right) = \beta_D : \Gamma_1 = \Gamma_0 \text{ for } T = 0 :$$
$$\Gamma_1 = \Gamma_0 + \beta_D \exp[-(U/T)].$$

(7)

Here the symbol $\Gamma_0$ stands for the linewidth of unbroadened line, while the symbol $w_D$ for the average jump frequency leading to another position of the resonant atom. The symbol $U$ denotes activation energy for the jump, as it is assumed that diffusion proceeds via classical over the barrier jump mechanism. Finally, the function $\rho(r)$ stands for the probability density function to jump distance $r$ from the origin during single uncorrelated jump. Parameters $\beta_D$ and $U$ could be treated usually as constants and they are adjustable parameters. Oscillatory and diffusive motions could be considered as uncorrelated one to another. The jump diffusion does not need to be truly long range provided the number of different accessible sites is very large, as the diffusive self-correlation function "remembers" usually only few past events.

Random jump-like rotation of the particle affects linewidth and quadrupole splitting provided it is much slower than the characteristic time scale of the quadrupole splitting. One can estimate splitting due to this motion as $\sqrt{\Delta^2 - 4w_Q^2} \approx \Delta \left[ 1 - 2(w_Q/\Delta)^2 \right]$. Here the symbol $\Delta$ stands for the static quadrupole splitting, while the symbol $w_Q \approx w_0 q^{-1} \int_\Omega d\Omega'\, \Omega'\, F(\Omega')$ describes random rotation dynamics. The symbol $w_0$ denotes average frequency of the uncorrelated rotational jumps, while the symbol $F(\Omega')$ denotes probability density function to rotate by the set of $\Omega'$ Eulerian angles during above jump. The integration goes over irreducible part of the properly extended unit sphere $\Omega$. For axially symmetric electric field gradient only polar angle is meaningful, otherwise all three Eulerian angles have some meaning. Above approximation is valid in the slow rotation limit. Otherwise, the full super-operator formalism has to be used[27-29]. Rotations are correlated with diffusive motion.

*3.2.2. Magnetic relaxation*

Slow dynamics occurs at the time scales longer than 100 ps extending to about 1 μs. One can assume again that conditions of thermal equilibrium hold. Another dynamical effect is relaxation of the hyperfine magnetic field to null due to the thermal fluctuations of the magnetic moments within nanoparticle. A blocking temperature is well defined for a particular nanoparticle and critically depends upon the size of the particle. Isolated particles practically do not interact one with another particularly having antiferromagnetic internal order. However, even partial removal of the embedding matrix leads to the clustering of particles and moves blocking temperatures up. Internal thermal disorder of the magnetic moments occurs as well and may depend on the particle size. The latter disorder belongs to the realm of the fast dynamics. Above comments apply to the particles with internal magnetic order[30].

## 4. Results

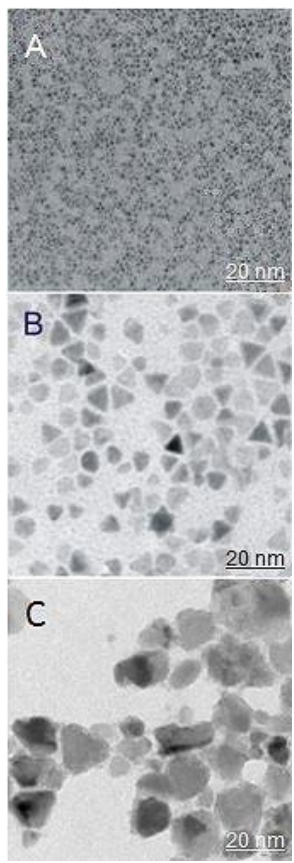

Some images obtained by transmission electron microscopy (TEM) are shown in Figure 1.

The images clearly confirm that the particles of 3 nm, 9 nm and 20 nm are separated one from each other giving a chance for higher level dynamics to occur. The detailed characterization of the nanocrystals by X-ray diffraction and energy-dispersive X-ray spectroscopy can be found in the Supporting Information (Figure S1 and S2). The average crystallite sizes, determined from the Scherrer formula, are 4.3, 11.6 and 16.3 nm for the A, B and C samples, respectively (Table S2, Supporting Information) in a good agreement with the TEM images.

Figure 1 Transmission electron microscopy images of ~3 nm (A), ~9 nm (B), and ~20 nm (C) particles with organic ligands. See some higher order structures formed by nanoparticles – particularly by ~9 nm particles having regular pyramidal shape. See also separation of particles.

Mössbauer spectra obtained for 20 nm particles stabilized by organic ligands are shown in Figure 2 for a temperature range 80 – 240 K. They were fitted within standard transmission integral approach. One can see that the hyperfine magnetic field is already averaged to null due to the fast magnetic relaxation in the whole temperature range. The spectrum could be fitted by some distribution of the quadrupole split doublets due to the intrinsic defects of the particles. Actually the absorption cross-section is described quite satisfactorily by four symmetrical doublets having individually fitted splitting, total shift and contributions. Linewidth could be set as common for all doublets. Contributions obtained at 80 K could be fixed at the same values for higher temperatures. All four components are characterized by small linewidth and thus called as narrow component ($C_1$). Broad components ($C_2$) appear at about 200 K. Note dramatic drop of the recoilless fraction with increasing temperature accompanied by the line broadening. Figure 3 shows spectra of 3 nm particles stabilized by organic ligands and obtained at various temperatures.

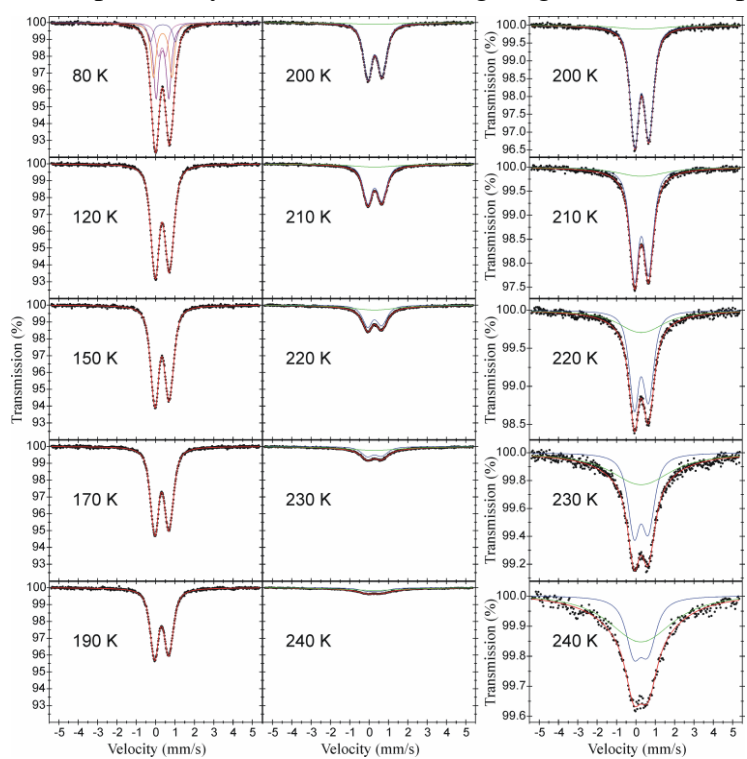

**Figure 2** Spectra of the 20 nm particles stabilized by organic ligands. Individual doublets are shown for 80 K spectrum. The average broad and narrow components are shown since 200 K. Right column shows spectra since 200 K with expanded vertical scale.

**Figure 3** Spectra of 3 nm particles stabilized by organic ligands and obtained at various temperatures. The average broad and narrow components are shown for 230 K spectrum. Inset (230 K) shows spectrum with expanded vertical scale. Note magnetic hyperfine interaction at 6.2 K. The magnetic splitting is due to the fact that magnetic relaxation is blocked at such low temperature like 6.2 K.

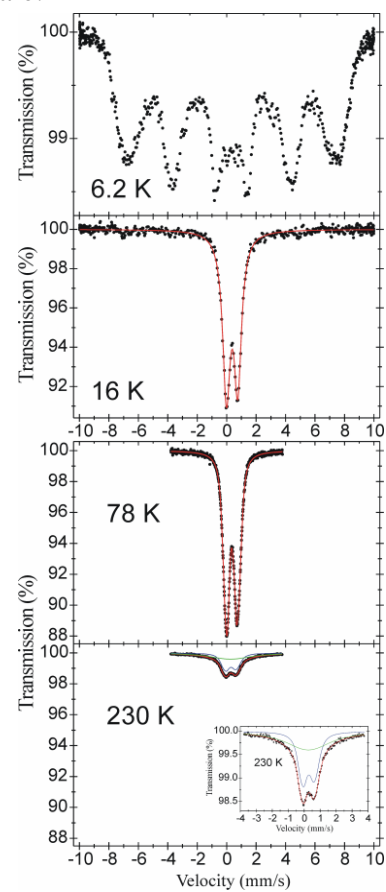

## 5. Discussion

*5.1. Fast dynamics*

Figure 4 shows weighted average total spectral shift $S$ versus shift in room temperature α-Fe plotted against temperature for 20 nm and 3 nm particles with organic ligands. It is assumed that variation of the isomer shifts versus temperature is negligible in the investigated temperature range.

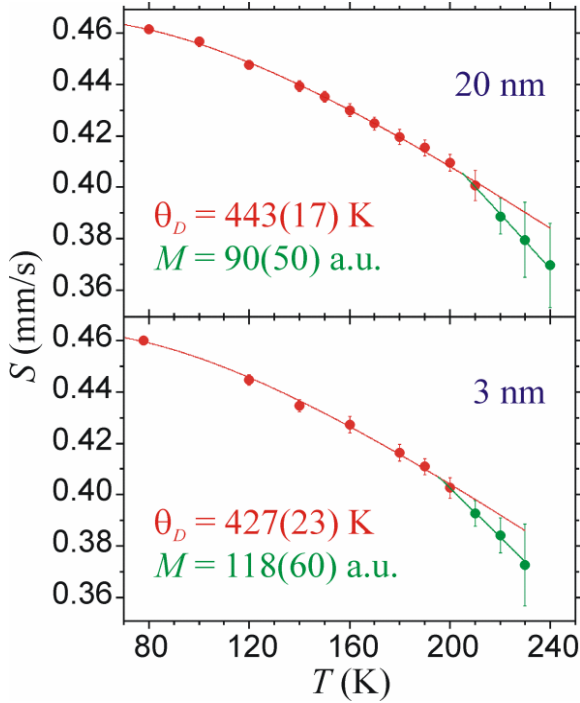

**Figure 4** Total spectral shift $S$ versus room temperature α-Fe plotted versus temperature for 20 nm and 3 nm particles with organic ligands. Points marked in red were used to obtain Debye temperatures, while points marked in green were used to obtain additional shift due to the internal vibrations of the particle. The latter shift yields effective mass for additional vibrations.

One can note that Debye temperatures for both kinds of particles are practically the same within the experimental error. Errors on temperatures $T_D$ are too big to draw any conclusions. It is interesting to note that Debye temperature estimated for the bulk chalcopyrite Cu-Fe-$S_2$ is definitely lower than observed by us for $^{57}$Fe in nanoparticles [31]. One has to observe that Debye temperature satisfies the relationship $\theta_D = \frac{4}{3}\int_0^{+\infty} dx\, x\, D(x)$ due to the normalization condition. Hence, increase of the Debye temperature might be caused by the lack of the low frequency modes, the latter being absent due to the effect of confinement. Slightly smaller total shift for 3 nm particles at 80 K in comparison with 20 nm particles at the same temperature might be an indication for increase of the electron density on the iron nucleus with decreasing size of the particle.

Interaction of the particle with the thermal bath provided by the network of organic ligands induces oscillations within particle leading to the additional SOD at elevated temperature. Such effect has not been observed previously to our best knowledge despite intensive search for[32]. The points above 200 K, marked in green in Figure 4, were used to obtain this additional shift quantitatively. The shift yields effective mass for additional vibrations. (see, Section 3.1.1). The statement on too big errors to draw any conclusions applies also to the effective masses, albeit larger mass for smaller particles might be an indication that it is harder to excite internal vibrations in a smaller particle as it is more rigid in comparison with larger particle. Accidentally the effective mass for 20 nm particles is very close to the mass of the Fe-S complex. It is likely that FePt particles with similar organic ligands behave in a similar way[13,14]. On the other hand, bulk material does not show any such vibrations due to the negligible surface to volume ratio.

The ratio of the recoilless fractions $f/f_0$ is shown in the upper part of Figure 5 for 20 nm particles with organic ligands. A reference temperature is chosen as the lowest temperature of the series, i.e., as $T_0 = 80\,\text{K}$. Inset shows relative spectral area (RSA) calculated as[19]:

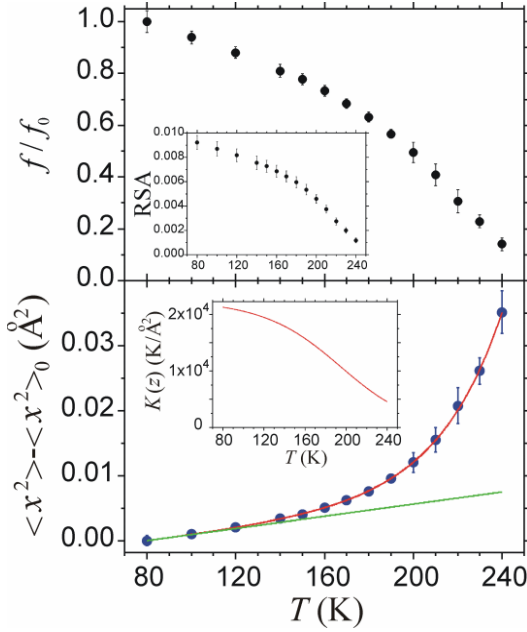

$$\text{RSA} = N_C^{-1} \sum_{n=1}^{N_C} \left( \frac{N_0 - N_n}{N_0} \right).$$

(8)

Here the symbol $N_C$ stands for the number of channels in a folded spectrum, the symbol $N_0$ denotes baseline, i.e., the number of counts per data channel far off the resonance, while the symbol $N_n$ stands for the number of counts in a particular $n$-th data channel.

**Figure 5** Upper part shows relative recoilless fraction $f/f_0$ plotted versus temperature for 20 nm particles with organic ligands. Inset shows relative spectral area (RSA) versus temperature – see text for details. Lower part shows function $\langle x^2 \rangle - \langle x^2 \rangle_0$ plotted versus temperature for 20 nm particles with organic ligands. The straight green line corresponds to the constant spring constant, i.e., to $\beta = 0$ condition. Inset shows spring constant $K(z)$ versus temperature with $z = T - T_0$. A Debye temperature was set to $\theta_F = 443\,\text{K}$.

Actually shape of RSA and shape of the ratio $f/f_0$ are very similar as the sample is resonantly thin at all investigated temperatures. Corresponding function $\langle x^2 \rangle - \langle x^2 \rangle_0$ is shown in the lower part of Figure 5. It was calculated by setting $\theta_D \approx \theta_F = 443\,\text{K}$, the latter value being obtained from SOD as shown in Figure 4. The following parameters were obtained from the fit to the expression (3) for 20 nm particles with organic ligands: $\alpha = 4.7(7) \times 10^{-5}$ (Å²/K), $\beta = 1.3(7) \times 10^{-7}$ (Å²/K²), and $\varepsilon = 3.5(6) \times 10^{-2}$ (K⁻¹). Spring constant $K(z)$ shown in the inset of the lower part of Figure 5 is obtained from the expression (4). Its fast decreasing with increasing temperatures shows that motion of the particles could be described by the effective classical oscillator with the spring constant dramatically softening with the increasing temperature.

*5.2. Slow dynamics*

Relative contributions of the narrow component $C_1 = 1 - C_2$ and broad component $C_2$ are shown versus temperature in the uppermost part of Figure 6 for 20 nm particles with organic ligands.

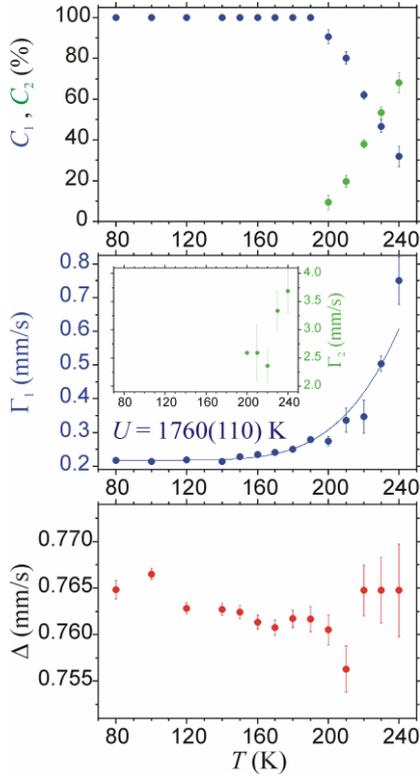

**Figure 6** The uppermost part shows relative contributions of the narrow component $C_1 = 1 - C_2$ and broad component $C_2$ plotted versus temperature for 20 nm particles with organic ligands. The central part shows linewidth $\Gamma_1$ plotted versus temperature for 20 nm particles with organic ligands. The activation energy for the long range like isotropic jump-like diffusion amounts to $U = 1760(110)$ K. Inset shows $\Gamma_2$ versus temperature. The fit at 200 K was performed with fixed value of $\Gamma_2$ at value obtained for 210 K and therefore it has no error bar. The lowest part shows weighted average of the quadrupole splitting $\Delta$ plotted versus temperature for 20 nm particles with organic ligands.

Slow dynamics is characterized by the motion described by the overdamped classical oscillator model. In addition activated diffusive motion is seen and it has long range like character. It is not true long range diffusion as some still higher level order in the form of Abrikosov vortices forming highly disordered Abrikosov lattice[33] survives on the long time scale till approximately room temperature as seen in Figure 1. Central part of Figure 6 shows linewidths $\Gamma_1$ and $\Gamma_2$ (corresponding to $C_1$ and $C_2$ components, respectively) versus temperature for 20 nm particles with organic ligands. As it is found from the fit of $\Gamma_1$ to the formula (7), the activation energy for the long range like isotropic jump-like diffusion amounts to $U = 1760(110)$ K.

Lower part of Figure 6 shows weighted average of the quadrupole splitting $\Delta$ versus temperature for 20 nm particles with organic ligands. Since quadrupole splitting remains independent of temperature, it is clear that none significant rotations occur on the time scale experimentally accessible and within accessible temperature range (jump-like rotation should affect both linewidth and quadrupole splitting as discussed in sub-section 3.2.2). This finding is in agreement with the image of the part B of Figure 1. One can clearly see higher level order of particles within part B of Figure 1. Such order prevents rapid rotations as particles are highly non-spherical and relatively close one to another within observed vortices. One can assume as well that dynamics is similar within Mössbauer absorber and specimen used in electron microscopy. Sufficiently fast rotations with sufficiently large rotation angles leading to decrease of the observed electric field gradient would smear image to nearly spherical appearance of the particles.

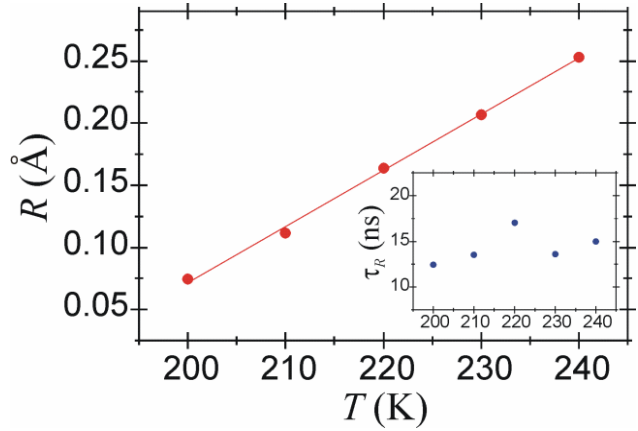

**Figure 7** "Radius" of the overdamped oscillator $R$ plotted versus temperature for 20 nm particles with organic ligands. Corresponding time scale $\tau_R$ is shown versus temperature in the inset. Error bars are not calculated for $R$ and $\tau_R$ parameters.

The "radius" $R$ and time scale $\tau_R$ are shown in Figure 7 for 20 nm particles without error bars as accuracy of the experimental data (particularly for $\Gamma_2$) allows to show only general trend versus temperature. They follow expressions (5) and (6). Rapid and fairly linear increase of the "radius" $R$ with increasing temperature (~0.0045 Å/K) indicates gradual "melting" of the matrix. Hence, particles move in the "cavities" with gradually increased size with increasing temperature, and the super-lattice composed of particles exhibits strongly non-linear behavior at elevated temperatures.

Finally, as seen from Figure 3, a magnetic dipole hyperfine interaction is visible only at 6.2 K and some magnetically induced broadening could be still seen at 16 K for 3 nm particles. This confirms that our nanoparticles are small enough for their blocking temperature (superparamagnetic effect) to be quite low. Only at so low temperature the spin relaxation is slow enough to result in the non-zero magnetic hyperfine field during observation time[30]. However, from the Mössbauer spectra measured without external magnetic field it is impossible to distinguish whether the particles are ferro- or antiferromagnetic. The iron magnetic moments in the bulk α-chalcopyrite (tetragonal) order antiferromagnetically well above room temperature[34-36].

## 6. Conclusions

Fast dynamics exhibits two levels. One is practically the same as for the bulk material except Debye temperature, while the other one (higher) could be described by the effective classical oscillator with the spring constant dramatically softening with the increasing temperature. Actually Debye temperature for the iron atom within particle is higher than estimated for the bulk material. This effect is likely to be due to the extinction of the low frequency acoustic phonon modes within nanoparticles. Interaction of the particle with the thermal bath provided by the network of organic ligands induces oscillations within particle leading to the additional SOD at elevated temperature.

Slow dynamics is characterized by the motion described by the overdamped classical oscillator model. In addition activated diffusive motion is seen and it has long range like character. No measurable rotations of the particle were found in the accessible temperature range.

Finally, isolated nanoparticles of the chalcopyrite stabilized by organic ligands exhibit rapid relaxation of the hyperfine magnetic field. The relaxation vanishes for the smallest particles at about 12 K. It is rather unlikely that the magnetic ordering temperature is lowered from well above room temperature to about 12 K by defects induced by the dispersing bulk material into small particles possessing still quite rigid structure.

## Acknowledgments

We are grateful to Prof. S. M. Dubiel, AGH University of Science and Technology, Kraków, Poland for providing us with his spectrometer for Mössbauer measurements at $T = 6.2\,\text{K}$ and $16\,\text{K}$. G.G., P.B. and A.P. wish to acknowledge financial support from the National Centre of Science in Poland (NCN, Grant No. 2015/17/B/ST4/03837).

TOC Graphics

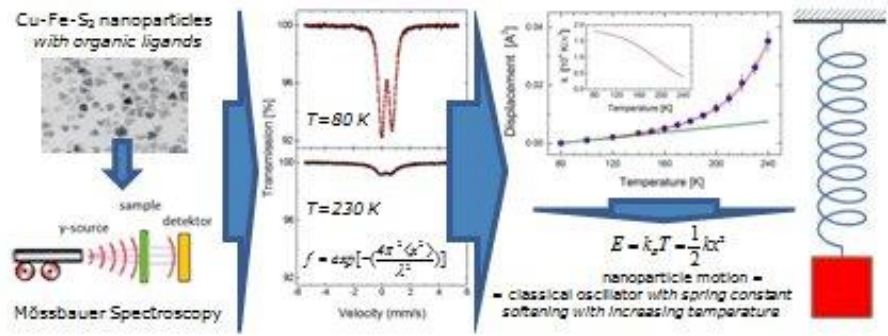

**Supporting Information**

**Dynamics of Ternary Cu-Fe-S$_2$ Nanoparticles Stabilized by Organic Ligands**

J. Żukrowski[1], A. Błachowski[2], K. Komędera[2], K. Ruebenbauer[2], G. Gabka[3], P. Bujak[3], A. Pron[3], and M. Przybylski[1,4*]

[1]Academic Centre for Materials and Nanotechnology, AGH University of Science and Technology
*PL-30-059 Kraków, Av. A. Mickiewicza 30, Poland*

[2]Mössbauer Spectroscopy Division, Institute of Physics, Pedagogical University
*PL-30-084 Kraków, ul. Podchorążych 2, Poland*

[3]Faculty of Chemistry, Warsaw University of Technology
*PL-00-664 Warsaw, ul. Noakowskiego 3, Poland*

[4]Faculty of Physics and Applied Computer Science, AGH University of Science and Technology
*PL-30-059 Kraków, Av. A. Mickiewicza 30, Poland*

[*]**Corresponding author:** marprzyb@agh.edu.pl

**PACS:** 05.40.-a; 64.70.pv; 66.30.Pa; 76.80.+y

**Short title:** Dynamics of nanoparticles with organic ligands

Experimental

X-ray powder diffractograms were recorded at room temperature on a Bruker D8 Advance diffractometer equipped with a LYNXEYE position-sensitive detector using Cu Kα radiation ($\lambda = 0.15418$ nm). The data were collected in the Bragg-Brentano (θ/2θ) horizontal geometry (flat reflection mode) between 10° and 70° (2θ) in a continuous scan, using 0.04° steps at 10 s/step. The incident-beam path in the diffractometer was equipped with a 2.5° Soller slit and a 1.14° fixed divergence slit, whereas the path of the diffracted beam was equipped with a programmable antiscatter slit (fixed at 2.20°), a Ni filter to remove Cu-β radiation, and a 2.5° Soller slit. The sample holder was rotated at an angular speed of 15 rpm. The data were collected under standard conditions (temperature and relative humidity). Transmission electron microscopy (TEM) analysis was performed on a Zeiss Libra 120 electron microscope operating at 120 kV. Elemental analysis was carried out with a multichannel Quantax 400 energy-dispersive X-ray spectroscopy (EDS) system with a 125 eV resolution xFlash Detector 5010 (Bruker) using a 15 kV electron beam energy.

**Table S1.** Summary of the synthesis conditions and composition of Cu-Fe-S nanocrystals.

|   | CuCl (mg) | FeCl$_3$ (mg) | thiourea (mg) | OA[a] (mg) | Cu/Fe/S/OA[b] | Cu/Fe/S[c] | Size (nm) |
|---|---|---|---|---|---|---|---|
| **A** | 60 | 100 | 93 | 361 | **1.0/1.0/2.0/2.1** | **1.62/1.00/2.01** | 2.7 ± 0.3 |
| **B** | 60 | 100 | 93 | 207 | **1.0/1.0/2.0/1.2** | **1.64/1.00/2.04** | 9.3 ± 1.7 |
| **C** | 60 | 100 | 93 | 100 | **1.0/1.0/2.0/0.6** | **1.76/1.00/2.04** | 19.2 ± 2.4 |

[a]oleic acid; [b]molar ratio of precursors; [c]ratio of elements in the nanocrystals from EDS; [d]size of the Cu-Fe-S nanocrystals determined from TEM images.

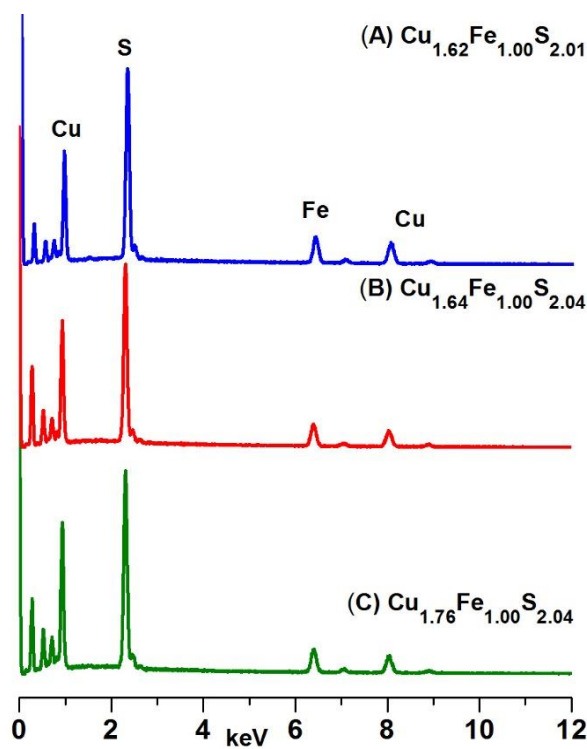

**Figure S1.** Energy-dispersive spectra of Cu-Fe-S nanocrystals of batches **A-C**.

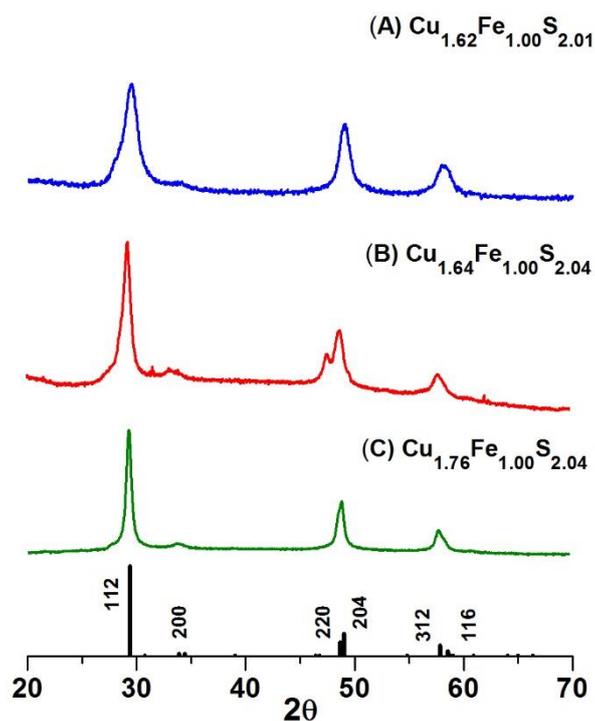

**Figure S2.** The X-ray diffractograms of the Cu-Fe-S nanocrystals obtained for a fixed precursor ratio (Cu:Fe:S = 1:1:2) and varying ligand (OA) to precursor ratios: OA:Cu = 2.1 (**A**); OA:Cu = 1.2 (**B**); OA:Cu = 0.6 (**C**). The black bars indicate the diffraction pattern of chalcopyrite $CuFeS_2$ (JCPDS 37-0471).

**Table S2.** Average nanocrystalline size of Cu-Fe-S samples: A, B and C calculated based on X-ray diffractograms using Scherrer's formula: $FWHM(2\theta) = \dfrac{K \times \lambda}{L \times \cos\theta}$, K = 1.00, λ = 1.5418 Å.

| Sample | $2\theta$ (deg) | FWHM (deg) | FWHM (rad) | L (nm) |
|--------|-----------------|------------|------------|--------|
| A      | 29.5            | 2.14       | 0.0374     | 4.3    |
| B      | 29.3            | 0.78       | 0.0137     | 11.6   |
| C      | 29.3            | 0.56       | 0.0098     | 16.3   |